\documentclass[12pt,a4paper]{article}
\usepackage{graphicx}
\usepackage{amsmath}
\usepackage{amssymb}
\usepackage{cite}
\usepackage[height=8.8in,width=6.45in]{geometry}

\numberwithin{equation}{section}

\begin{document}
\begin{titlepage}
\hfill CALT-TH 2016-027, IPMU16-0139\\
\vbox{
    \halign{#\hfil         \cr
           } 
      }  
\vspace*{15mm}
\begin{center}
{\Large \bf Non-supersymmetric AdS and the Swampland}

\vspace*{15mm}

{\large Hirosi Ooguri$^{a,b}$ and Cumrun Vafa$^c$}
\vspace*{8mm}

${}^a$ Walter Burke Institute for Theoretical Physics \\ California
Institute of Technology,
 Pasadena, CA 91125, USA\\
\vspace*{0.2cm}
${}^b$ Kavli Institute for the Physics and Mathematics of the Universe
\\ University of Tokyo,
Kashiwa, 277-8583, Japan\\
\vspace*{0.2cm}
${}^c$
Center for the Fundamental Laws of Nature \\
Harvard University, Cambridge, MA 02138, USA\\

\vspace*{0.7cm}

\end{center}
\begin{abstract}
We propose to sharpen the weak gravity
conjecture by the statement
that, except for BPS states in a supersymmetric theory, the gravitational force is strictly weaker than any electric force
and provide a number of evidences for this statement.
Our conjecture
implies that any
non-supersymmetric anti-de Sitter vacuum
supported by fluxes must be unstable, as is
the case for all known attempts at such holographic constructions.

\end{abstract}

\end{titlepage}

\vskip 1cm

\section{Introduction}
The weak gravity conjecture (WGC) \cite{ArkaniHamed:2006dz}
is an example of how seemingly consistent low energy gravitational theories can fail to have ultra-violet completion
and instead belong to the swampland \cite{Vafa:2005ui}; see
\cite{Ooguri:2006in} for more swampland constraints.
Recently, there have
been a number of papers checking and extending the WGC
and applying it to cosmology \cite{delaFuente:2014aca,Rudelius:2015xta,Bachlechner:2015qja,Heidenreich:2015wga,Brown:2015iha,Ibanez:2015fcv, Hebecker:2015zss, OIST}.
The conjecture has also been related to the holographic principle in
\cite{Harlow:2015lma}.
In this brief note, we propose sharpening of the WGC and point out that
it implies that
non-supersymmetric anti-de Sitter (AdS) vacua supported by fluxes are unstable.

The WGC states that the mass of a particle carrying a basic unit
of charge or its small multiple is less than or equal to that predicted
if the particle were an extremal black hole.  In other words, the
gravitational force
is equal or weaker than the electric force between such particles.
 The same idea applies to charged branes, where the tension
for a basic brane charge or its small multiple should be less than or
equal to
that of the corresponding extremal black brane.
The fact that it can sometimes be equal is
 well-known in string theory:  BPS states in a supersymmetric theory
saturate this relation.  When the equality holds,
 the gravitational attraction and the electric
repulsion of branes cancel out exactly.

In this paper we conjecture that the inequality between gravitational and
electric
forces is saturated if and only if the underlying theory is supersymmetric
and the states under consideration are BPS with respect to the
supersymmetry. We study a number of examples and provide evidences for this new conjecture.

This may seem like an innocuous extension of the WGC,
but it has dramatic consequences. In particular,  it implies that
non-supersymmetric AdS vacua
supported by fluxes\footnote{By this we mean that there is a $(p+1)$-form
gauge potential whose field strength along AdS does not vanish and is given by 
$F_{\mu_1,...,\mu_{p+2}}=N\epsilon_{\mu_1,...,\mu_{p+2}}^{{\rm AdS}}$.} must be unstable and
that their effective theories belong to the swampland, even if they may look consistent.   
 In all known top-down constructions from M/string theory, AdS geometry is supported
by some flux. The sharpened version of the WGC predicts that
there is a brane charged with respect to the flux and with tension
less than the charge. 
As shown in \cite{Maldacena:1998uz} and reviewed
in section 3 of this paper, such branes can nucleate in AdS,
and the Coulomb repulsion by the flux wins over the tension 
to expand the brane. It then reaches the boundary of AdS within 
finite time, reducing the flux. 

We can also understand this instability
in the following way.
To construct AdS holography in M/string theory, we
typically start with a large number of extremal branes and 
take its near horizon limit. The sharpened version of the 
WGC implies that, unless the branes are BPS states of a supersymmetric theory, their electric repulsion wins over their gravitational attraction and this system is unstable.  
In the gravitational dual description, which will presumably exist when the extremal brane is large, the decay process can be slow
and it can exist as a quasi-stable state.  
However, due to the gravitational time delay effect,
the lifetime becomes shorter as we measure it closer to the horizon.
In the near horizon limit, the decay becomes instantaneous.  Therefore holographic duality may continue to exist for non-supersymmetric 
system, but its near horizon geometry, which leads to AdS dual description, does not exist for the non-supersymmetric system because of this instability.

The lack of non-supersymmetric AdS dual description, motivates us to propose our second conjecture, which is basically removing the assumption that the AdS is supported by the flux:  We conjecture that 
non-supersymmetric
AdS holography is not realizable in a consistent quantum theory with 
low energy description in term of the Einstein gravity coupled 
to a finite number of matter fields, even if the AdS is not supported by a flux.  
We are restricting to theories
with the Einstein gravity description in low energy 
since the WGC is motivated by black hole physics.\footnote{Therefore,
topological field theories such as the Chern-Simons gauge theory
and the Donaldson theory (though they have diffeomorphism invariance) are excluded
from the scope of this conjecture. }  Note that this second conjecture does not follow from the sharpened version of the WGC and it is conceivable that the sharpened version of WGC is correct but this second conjecture is false.

Note that our second conjecture would have counter-examples if we relax the condition that the number of matter fields is finite.
For example, non-supersymmetric higher spin theories of the Vasiliev-type
may be quantized consistently in AdS, and some of them may be dual to
non-supersymmetric conformal field theories \cite{Giombi:2016ejx}.
However, these theories contain infinite towers of higher spin fields.
Another example is the Sachdev-Ye-Kitaev model \cite{Sachdev:1992fk,
  Sachdev:2010um, Kitaev}. Although this model saturates the chaos bound
  on the Lyapunov exponent \cite{Kitaev} and some of its correlation functions are
  well-approximated by the dilaton gravity in two dimensions first studied in 
  \cite{Jackiw, Teitelboim}, its spectrum contains a tower of states 
  with an approximately integer spacing, which can be interpreted as
  stringy states in the bulk whose tension is of the order of the AdS radius
\cite{Polchinski:2016xgd, Maldacena:2016hyu}.   Thus this theory is not a pure Einstein gravity coupled to a finite number of fields at the AdS scale.

The lack of explicit examples of non-supersymmetric AdS holography is
often attributed to technical difficulties in estimating non-perturbative 
effects. 
Our conjecture suggests that there is a more fundamental reason for the lack of
examples.  
If true, the absence of non-supersymmetric AdS vacua
will have a significant implication on
applications of the AdS/CFT correspondence to condensed matter physics,
quantum information, hadron physics, and other areas in theoretical physics 
since interesting models in these contexts are typically those without supersymmetry. 

Conformal field theory dual to the Einstein gravity with a finite number of matter fields
must satisfy the gap condition: such a theory can contain only a small number of
primary fields whose operator products generate all primary fields up to a
certain energy scale, which can be made parametrically large in the large $N$ limit. 
The bulk reconstruction argument seems to suggest that the converse is also true
\cite{Heemskerk:2009pn, Heemskerk:2012mn}.
If we accept this premise, our conjecture would imply 
that non-supersymmetric conformal field theories cannot satisfy 
the gap condition. 

Our conjectures can have highly non-trivial consequences.  For example,
our second conjecture, combined with the observation in \cite{ArkaniHamed:2007gg} that the standard model with minimal Majorana neutrino masses compactified on a circle leads to non-supersymmetric AdS$_3$ vacua with large radius,\footnote{We are assuming that there is no hidden mechanism in the standard model destablizing the solutions found in \cite{ArkaniHamed:2007gg}.} rules out minimal Majorana neutrino masses for standard model to avoid this possibility!

The organization of this note is as follows.  In section 2, we discuss
several examples for the WGC and show how they satisfy the strict
inequality for non-BPS states.  In section 3 we motivate our conjecture
that non-supersymmetric AdS holography does not exist and discuss
several examples to support the conjecture.

\section{Evidence for Sharpened Version of WGC}
\bigskip
In the simplest form, the WGC states that, for a $U(1)$ gauge theory
coupled to gravity, there exists an elementary particle of mass $m$ and
charge $Q$ satisfying the 
inequality,
\begin{equation}
Gm^2={\left( {m\over M_{\rm{Pl}}}\right)}^2\leq Q^2.
\label{WGC}
\end{equation}
Here $G$ is the Newton constant and $M_{\rm{Pl}} =1/G$ is the Planck mass.
The charge  $Q$ is supposed to be the basic unit of the electric charge
or a small multiple of it. We 
assume $m<M_{\rm{Pl}}$ so that any charged black hole with 
masses greater than $M_{\rm{Pl}}$ and satisfying the 
Reissner-Nordstr\"om bound
is able to emit such a particle. We have suppressed
a numerical factor that depends on spacetime dimensions; the precise form of the
inequality is given later.
The reason for allowing equality in the above is that this can be saturated
in a supersymmetric theory with BPS states, which often
happens in string theory. Otherwise, allowing the equality above seems
unnatural for the following reasons.

One of the original motivations for the WGC in  \cite{ArkaniHamed:2006dz} was 
to allow extremal black holes to decay  unless protected by
supersymmetry since their existence would lead to
a large number of remnants. 
   If we interpret this requirement literarily, 
(\ref{WGC}) should not include the equality 
since extremal black holes can only emit particles strictly 
below the
Reissner-Nordstr\"om bound.
Particles at the bound  would have vanishing phase space for their emissions. 
Moreover,  such particles would be on the verge of violating the WGC!
If we allow the equality, there should be a good reason why small perturbation
to the theory would not tip the balance in a wrong direction and
violate the WGC.  In the supersymmetric case, the robustness is guaranteed by
the BPS condition.
For a non-supersymmetric theory, however, if the inequality is saturated,
there is no known mechanism to keep this balance from tipping in the wrong
direction. This motivates the stronger conjecture:
\bigskip

\noindent
\hskip 2cm
{\it The WGC is saturated if and only if the theory is supersymmetric}

\noindent
 \hskip 2cm {\it and
the state in question is a BPS state.}

\bigskip
\noindent
The same sharpening of the conjecture can be considered for branes also:
a consistent gravity theory with a $(p+1)$-form gauge field must contain
$p$-branes below the $p$-brane version of the Reissner-Nordstr\"om bound, unless the
$(p+1)$-form is in the supergravity multiplet.
Such $p$-branes would ensure that black $p$-branes are unstable unless
protected by supersymmetry. 

Note that the original
paper  \cite{ArkaniHamed:2006dz}  on the WGC does not claim that there is
a particle with $|m/Q| \leq  M_{\rm{Pl}}$ for every possible charge $Q$. 
In fact, the paper described counter-examples to it.  
The sharpened version of the
conjecture also applies only to particles
with minimal $|m/Q|$ and the statement is that there are some non-supersymmetric states for which
 this ratio is strictly less than $M_{\rm{Pl}}$.
The lowest $|m/Q|$ ratio may occur for a state whose charge is a small multiple 
of the basic charge. 

Let us present evidences to demonstrate our conjecture.
Suppose there is a family of stable non-BPS states which carry different values of $Q$
and become extremal black holes in the limit of large $Q$,
namely, $m(Q)/|Q| \rightarrow M_{\rm{Pl}} $ for $Q \rightarrow \infty$.
In \cite{lubos}, finite $Q$ corrections to $m(Q)/|Q|$ are computed
for non-supersymmetric black holes 
using higher derivative terms in the low energy 
effective action derived from the heterotic string theory compactified on a torus.
It turns out that leading corrections take the form,  
\begin{equation}
  \frac{ m(Q)}{M_{\rm{Pl}}} = |Q| \left( 1 - \varepsilon(Q)\right),
  \label{corrections}
  \end{equation}
where $\varepsilon(Q)$ is always positive. 
This shows that, as the charge $Q$ decreases, 
the extremal black hole becomes super-extremal due to 
 higher derivative corrections and the strict 
WGC inequality is satisfied.

There are several examples of stable non-supersymmetric states that 
satisfy the strict inequality. One class of such states has already been discussed in
the original paper \cite{ArkaniHamed:2006dz}.
Consider the Narain compactification of the heterotic string theory on $T^d$.
The low energy theory has
$U(1)^{16+2d}$ gauge symmetry, and their charges
 make an even self-dual lattice,
$$(p_L,p_R)\in \Gamma^{16+d,d}.$$
In our convention, the left-mover is bosonic and the right-mover is
supersymmetric. Let us look for
the lowest mass state for a given set of charges.
Recall that the mass squared of a perturbative string state is given by,
$$m^2={1\over 2} p_L^2+N_L-1={1\over 2}p_R^2+N_R,$$
where $N_L$ and $N_R$ are amounts of excitations in the left and
right-movers.
To minimize the mass, we should try to take $N_L$ and $N_R$ as small as
possible.
Since ${1 \over 2}\left( p_L^2 - p_R^2 \right) = N_R - N_L + 1$,
we can set both $N_L$ and $N_R$ to be zero only if $p_L^2 - p_R^2 = 2$.
If $p_L^2 -p_R^2 < 2$, we can set $N_R=0$, but
 $N_L$ has to be non-zero. The corresponding
 state is BPS and known as a Dabholkar-Harvey state
\cite{Dabholkar:1989jt, Dabholkar:1990yf}.  In that case, $m^2={1\over 2}
p_R^2$, saturating
 the WGC bound.  On the other hand, if $p_L^2 - p_R^2 > 2$, we can choose
$N_L=0$
 but  $N_R$ has to be non-zero. The resulting state is not BPS, and we have,
$$m^2={1 \over 2} p_L^2-1,$$
leading to the strict inequality $m^2 < {1 \over 2} p_L^2$ because of the
Casimir
energy $(-1)$ for the bosonic left-moving sector on the worldsheet.  We
thus see the clear correlation between breaking supersymmetry and
realizing the strict WGC bound in this example.
For large $p_L^2$, the above mass formula takes the form of 
(\ref{corrections}) with $\varepsilon \sim 1/p_L^2$, which is  positive and
vanishes in the large charge limit $p_L^2 \rightarrow \infty$ as expected
from the analysis of \cite{lubos}.

The above example is in the weak
coupling limit of the heterotic string theory.
Both the WGC and its sharpened version are supposed to hold  for
arbitrary coupling constant.
We can test the sharpened version in the strong coupling limit
of the heterotic string theory compactified on $S^1$ by using its duality to
the Type I string theory \cite{Witten:1995ex, Polchinski:1995df}.
The low energy dynamics of the Type I D-string with the winding number $n$ on $S^1$
 can be described by an $O(n)$ gauge theory on the D-string worldsheet.
In particular, when $n=1$, the worldsheet
theory on the D-string is identical
to that on the fundamental heterotic string on $S^1$, except that the tension and the radius of $S^1$ are different
and relatd to those on the heterotic string side by the duality rules.
For example, the
 $O(1) =\mathbb{Z}_2$ gauge symmetry on the D-string worldsheet acts as the GSO projection
 on the heterotic string side. The spectrum on the singly-wound D-string therefore has the same structure
 as that on the heterotic string worldsheet discussed in the above,
 and  non-BPS excitations of the D-string satisfy the strict WGC inequality.
Thus, the sharpened version of the WGC 
works for both weak and strong coupling constants of heterotic string in this case.   It would be interesting to
generalize this analysis for $n > 1$.

Our conjecture also leads to a number of mathematically falsifiable predictions. 
Consider 
   M theory on K3, which is dual to the
heterotic string on $T^3$. Since the large volume limit of K3
in M theory
corresponds to the strong coupling limit in the heterotic string,
K3 geometry can be used to probe the regime
that is not accessible by the heterotic string perturbation.
This is specially so for non-supersymmetric states,
whose masses are not protected as we change the coupling constant.  Can
non-supersymmetric states have masses less than their charge in the large
volume limit of K3
on the M-theory side?
Consider $H_2(K3,\bf Z)$.  This gives a lattice with inner product, obtained by intersection of a pair of 2-cycles.  It is known that this lattice has signature $(19,3)$.  Moreover the metric on $K3$ induces a projection on 2-forms to self-dual and anti-self-dual forms.  This is called a polarization on $H^2(K3,{\bf Z})$ which induces a polarization on the dual space $H_2(K3,{\bf Z})$.  Using this we can identify $H_2(K3,{\bf Z})$ with the corresponding lattice of momenta $(p_L,p_R)$ for heterotic strings on $T^3$.  
When  $p_L^2 - p_R^2 \leq 2$, the $2$-cycle class can
be realized as a holomorphic curve of genus $g$, where
$$p_L^2-p_R^2=2-2g.$$
M2 branes wrapping holomorphic curves lead to BPS states,
and they are dual to the heterotic string BPS states.
On the other hand, if $p_L^2-p_R^2 > 2$, there is no holomorphic curve.
We can compare the minimum mass of the M2 brane
wrapping such a 2-cycle to its charges. In the large volume limit the
mass of such an M2 brane is equal to the area of the 2-cycle it
wraps. 
In particular suppose we have a polarization such that a charge vector associated to a 2-cycle is of the form $(p_L,0)$ with $p_L^2>2$.  Then this class cannot be represented by a holomorphic cycle.  On the other hand, suppose we have a minimal surface in this class.  Let $A$ denote the area of this cycle (where the volume of $K3$ is normalized so that for supersymmetric cycles  $A=|p_R|$).  Then the WGC implies
$$A< |p_L|,$$   
which would be interesting to verify.

There are other predictions that follow from the sharpened version of the WGC. 
For example, consider ${\cal N}=1$ supersymmetric theories in four dimensions
coupled to supergravity.
If the theory has $U(1)$ gauge symmetries,
the sharpened version of WGC predicts that there is a charged
state with mass strictly less than charge for small multiples of basic $U(1)$ charge.
This is because there are no BPS states in ${\cal N}=1$
theories in four dimensions.  For example, consider M theory
compactified on a $G_2$ holonomy manifold,
leading to ${\cal N}=1$ supersymmetric theory in four dimensions.  For each
2-cycle in the manifold, we get a $U(1)$ gauge factor.
M2 branes wrapping
the corresponding 2-cycles gives charged states.  The sharpened version of
the WGC predicts that the mass of the
corresponding M2 brane
is less than the corresponding charge.

Our final evidence  is provided
by
the behavior of the WGC inequality under dimensional reduction.
In \cite{Heidenreich:2015nta}, the precise form of the inequality is
determined
for a charged particle
with mass $m$ and charge $Q$ coupled to the Einstein-Maxwell-dilaton
system in $d$ dimensions as,
\begin{equation}
 8 \pi G\left( {\alpha^2 \over 2} + {d-3 \over d-2}
\right)m^2 \leq Q^2,
\label{WGCwithD}
\end{equation}
where $\alpha$ is the dilaton coupling to the Maxwell term as
$e^{-\alpha \phi} F^2$ in the Einstein frame for the metric.
If the dilaton is massless, it affects long range features of
black hole solutions and contributes to the WGC inequality.
Since the factor $(d-3)/(d-2)$ is an increasing function of $d$,
if the dilaton coupling $\alpha$ is independent of $d$,
a particle saturating the WGC inequality in $d$ dimensions 
would obey the strict WGC inequality
under toroidal compactification to lower dimensions.
However, the value of $\alpha$ can change
under compactification since
additional massless scalar fields such as the radion for the size of the
internal torus can be generated.
They can mix with the dilaton in the original theory
and modify the effective value of $\alpha$.
Remarkably, it was shown in \cite{Heidenreich:2015nta} that, if
compactification
preserves supersymmetry, the value of $\alpha$ is modified exactly
in such a way to compensate for the $d$ dependence in
 $(d-3)/(d-2)$ and to make the WGC inequality (\ref{WGCwithD})
 invariant under dimensional reduction. With supersymmetry, the dilaton
and the radion remain
 strictly massless in infrared. If supersymmetry is broken,
 scalar fields generically become massive and cease to contribute
 to long range features of black hole solutions.
 In such a case,  $\alpha$ cannot compensate
 for the $d$ dependence in in (\ref{WGC}),  and the inequality cannot be
saturated after
 dimensional reduction if it holds in the original theory.
 This gives yet another evidence for
 the sharpened version of the WGC, $i.e.$,
 the inequality cannot be saturated without supersymmetry.

\section{Gravitational Instability of Non-supersymmetric AdS}

The sharpened version of the WGC
implies that  any non-supersymmetric AdS geometry supported by flux is unstable
in a consistent quantum theory of gravity 
with low energy description in terms of the Einstein
gravity coupled to a finite number of matter fields. Equivalently,
there are no non-supersymmetric conformal field theories whose holographic
dual have such gravity description.
These statements follow from our sharpened version of the WGC combined with the analysis in \cite{Maldacena:1998uz} that
AdS$_{p+2}$ geometry supported by a $(p+2)$ form flux is unstable if
there are $p$-branes charged with respect to the flux and whose charges
are less than their tensions.  In the Euclidean AdS$_{p+2}$ geometry, 
$$ ds^2 = \cosh^2 \rho\ d\tau^2 + d\rho^2 + \sinh^2 \rho 
\ d\Omega_p^2 ,$$
 a spherical $p$-brane on $S^p$ at radius 
$\rho(\tau)$ evolving in the Euclidean time $\tau$ has the action
proportional to, 
$$ S_E \sim \int\left( \sinh^p \rho\  \sqrt{ \cosh^2 \rho + (d\rho/d\tau)^2 } - q \cdot \sinh^{p+1} \rho \right) d\tau , $$
where $q$ is the charge/tension ratio normalized in such a way that the Reissner-Nordstr\"om bound is at $q=1$.  
It was shown in  \cite{Maldacena:1998uz}  that,  when $q > 1$, there is an instanton solution
which nucleates a spherical brane with radiu $\rho_{\rm{n}}$ given by
$\tanh \rho_{\rm{n}} = 1/q$. The brane then expands in the Lorentzian AdS
as $\cosh \rho = \cosh \rho_{\rm{n}} / \cos t$ and reaches the AdS boundary at $\rho = \infty$ 
in finite time of $t= \pi/2$, reducing the flux of the geometry.

Unlike flat
or de Sitter vacua, the instability of AdS vacua can
be detected instantaneously in their would-be dual
conformal field theories \cite{Horowitz:2007pr, Harlow:2010az}.
This is because any observer in AdS gets access to the entire Cauchy
surface within a finite amount of time. In particular, an observer at the
boundary
of AdS can get access to an infinite volume space near the boundary within an infinitesimal
amount of time.
Since the instability process discussed in the above causes a finite decay
probability
per unit volume of AdS, the decay would happen instantaneously when seen
from the
boundary. Therefore, the dual conformal field theory cannot exist even as
a meta-stable
state.

The instantaneous nature of the decay can also be explained 
as follows. Consider an extremal $p$-brane, whose near horizon
reproduces the AdS$_{p+2}$ geometry in the above. 
According to the sharpened version of the WGC,
if the extremal $p$-brane is not protected by supersymmetry, 
there is a super-extremal brane charged with respect to 
the $(p+2)$-form flux and with tension less than its charge. 
The extremal brane can decay by emitting such super-extremal 
branes. A large extremal brane can exist as a quasi-stable state
since its decay rate can be small. However, the rate becomes larger
as we measure it closer to the horizon. In the near horizon limit,
the decay becomes instantaneous, consistently with the conclusion of the
previous paragraph.

There have been several proposals for non-supersymmetric AdS vacua.
One can start with AdS$_5 \times S^5$, which is dual to the ${\cal N}=4$
supersymmetric Yang-Mills
theory in four dimensions, and divide the $S^5$ by a discrete subgroup
$\Gamma$ of its $SU(4)$ rotational symmetry \cite{Kachru:1998ys}.
Supersymmetry is completely broken if the orbifold group
$\Gamma$ does not fit within an $SU(3)$ subgroup of the $SU(4)$ symmetry.
 It turns out that its gauge
theory
is typically not conformal when $\Gamma$ breaks supersymmetry completely
since couplings for double-trace operators cannot be stabilized under the
renormalization group flow. It was shown in \cite{Dymarsky:2005uh, Dymarsky:2005nc}
that, when the action of $\Gamma$ on $S^5$
has
a fixed point or when $\Gamma$ acts freely but the `t Hooft coupling is
small, a double-trace operator fails to find a fixed point. In this case, 
a closed string tachyon emerges in a twisted sector, and
the AdS$_5 \times S^5/\Gamma$ geometry becomes perturbatively
unstable.

When $\Gamma$ acts freely on $S^5$ and the `t Hooft coupling is large,
there is no tachyon in the spectrum. However, 
the geometry is still unstable non-perturbatively.
If $\Gamma$ acts freely on
$S^5$, there are non-contractible cycles on $S^5/\Gamma$. It was shown
in
\cite{Horowitz:2007pr} that there is an instanton solution which describe a tunneling of
the AdS$_5 \times S^5/\Gamma$ geometry into nothing.
In the instanton geometry, one of the supersymmetry breaking cycle shrinks to a point
and the geometry terminates over 
an $S^4$ homologous to the $S^4$ at the infinity of AdS$_5$. 
This is similar to the instanton 
used in \cite{Witten:1981gj} to demonstrate the instability of
the Kaluza-Klein vacuum. Here the role of the Kaluza-Klein circle
is played by the supersymmetry breaking cycle in $S^5/\Gamma$.
To conserve the 5-form flux in the geometry,
an appropriate number of D3 branes must wrap the $S^4$.
After analytical continuation to the Lorentzian signature,
the $S^4$ becomes an expanding bubble of $S^3$, which grows and 
reaches the boundary of AdS$_5$ in finite time, consuming the entire spacetime in the process.
The expanding bubble wrapped by D3 branes can be regarded an
example of
branes predicted by the weak gravity conjecture, except that its D3 charge is large. 
In this case, the brane carries exactly the amount of
charge
that cancels the 5-form flux on  $S^5/\Gamma$, rather than the minimum D3
brane charge.

There have been other proposals for non-supersymmetric AdS vacua, but
none of them has been demonstrated to be stable.
The non-supersymmetric AdS$_4$ vacua
in the massive IIA supergravity discussed in \cite{Narayan:2010em} allow
marginal decays in
their approximation, and the magnetic AdS geometry discussed in
\cite{Almuhairi:2011ws} does not stabilize the dilaton.

Since we know that there are conformal field theories without supersymmetry,
it is reasonable to ask: 
Why can't they have holographic AdS duals? To understand this, 
let us consider one more example where one attempts to construct 
a holographic dual to a non-supersymmetric conformal field theory.  
Consider the D1-D5 string on K3, whose low energy theory is described by the two-dimensional supersymmetric
sigma-model with the symmetric product of K3's as its target space. It is dual to 
the Type IIB string theory on AdS$_3 \times S^3 \times \ $ K3. 
However, the gravity approximation is not valid at the orbifold limit of the symmetric product
target space since the D1-D5 system there has more low energy states than predicted by the supergravity 
\cite{Dijkgraaf:1996xw, deBoer:1998us, Keller:2011xi, Hartman:2014oaa}.  We have to deform the sigma model away from the orbifold point by marginal operator corresponding to blow up modes where a pair of K3 approch one another.  This is needed in order to land in a regime where gravity dual does not have infinitely many light modes.
This shows that
 the existence of marginal operators of the symmetric product of the orbifold theory 
plays a key role in realizing the holographic dual described by the Einstein gravity with a finite number of matter fields.
 Now, let us contrast this with its non-supersymmetric counterpart. 
The bosonic sigma model on the symmetric product of K3's
is conformal at least in the orbifold limit.  
We can ask if it has a holographic dual.  
In the supersymmetric case, we saw that the existence of holographic dual is crucially hinged on the exactly marginal 
operators which can take the theory away from the orbifold limit.  
In the non-supersymmetric case, it is easy to check that the requisite blow-up mode is not marginal 
and that we cannot go away from the orbifold point.  This is consistent with our conjecture that 
this theory will fail to have a holographic dual as a weakly coupled gravity.

\section*{Acknowledgments}

We would like to thank N.~Arkani-Hamed, T.~Dumitrescu,
D.~Harlow, I.~Klebanov, J.~Maldacena, G.~Remmen, 
T.~Rudelius, A.~Sen, S.~Shenker, A.~Strominger, and E.~Witten for discussions.
The research of HO is supported in part by
U.S.\ Department of Energy grant DE-SC0011632, 
 by the Simons Investigator Award,
by the World Premier International Research Center Initiative,
MEXT, Japan,
by JSPS Grant-in-Aid for Scientific Research C-26400240,
and by JSPS Grant-in-Aid for Scientific Research on Innovative Areas
15H05895.
HO also thanks the hospitality of  
the Aspen Center for Physics, which is supported by
the National Science Foundation grant PHY-1066293,
and of the the Center for Mathematical
Sciences and Applications and the Center for the Fundamental Laws of Nature
at Harvard University.  The research of CV is supported by the NSF grant
PHY-1067976.

\end{document}